\documentclass[runningheads,a4paper,11pt]{llncs}

\usepackage[bookmarks, hidelinks]{hyperref} 
\usepackage{graphicx}
\usepackage{caption}
\usepackage{subcaption}
\usepackage{url}
\usepackage{xspace}

\newcommand{\seed}{\ensuremath{Seed}\xspace}
\newcommand{\salt}{\ensuremath{Salt}\xspace}
\newcommand{\passwordpolicy}{\ensuremath{PP}\xspace}

\newcommand{\accDataKey}{\ensuremath{K_{\mathrm{Data}}}\xspace}
\newcommand{\accDataKeyMac}{\ensuremath{K_{\mathrm{Data,Mac}}}\xspace}
\newcommand{\accDataKeyEnc}{\ensuremath{K_{\mathrm{Data,Enc}}}\xspace}
\newcommand{\accDataKeySSS}{\ensuremath{K_{\mathrm{Data,SSS}}}\xspace}
\newcommand{\accDataKeySSSGeneration}{\ensuremath{K_{\mathrm{Data,SSS}} = \mathrm{KDF}(K_{\mathrm{Data}}, url_{\mathrm{SSS}})}\xspace}
\newcommand{\accDataKeyMacSSS}{\ensuremath{K_{\mathrm{Data,SSS,Mac}}}\xspace}
\newcommand{\accDataKeyEncSSS}{\ensuremath{K_{\mathrm{Data,SSS,Enc}}}\xspace}

\newcommand{\secret}{\ensuremath{S}\xspace}
\newcommand{\secretDef}{\ensuremath{S = (\seed,\accDataKey)}\xspace}
\newcommand{\secretBD}{\ensuremath{S_{\mathrm{BD}}}\xspace}
\newcommand{\secretOTP}{\ensuremath{OTP_{\mathrm{BD}}}\xspace}
\newcommand{\secretOTPEnc}{\secretBD = \secret \ensuremath{\oplus} \secretOTP}
\newcommand{\secretOTPDec}{\secret = \secretBD \ensuremath{\oplus} \secretOTP}
\newcommand{\secretOTPUD}{\ensuremath{OTP_{\mathrm{UD}}}\xspace}
\newcommand{\secretOTPMask}{\ensuremath{M}\xspace}
\newcommand{\secretOTPMaskSSS}{\ensuremath{M_{\mathrm{SSS}}}\xspace}
\newcommand{\secretOTPMaskSSSGeneration}{\ensuremath{M_{\mathrm{SSS}} = \mathrm{PRG}(M, url_{\mathrm{SSS}}) }\xspace}
\newcommand{\secretOTPSSS}{\ensuremath{OTP_{\mathrm{UD,SSS}}}\xspace}
\newcommand{\secretOTPSSSGeneration}{\secretOTPSSS = \secretOTP \ensuremath{\oplus} \secretOTPMaskSSS}

\newcommand{\acc}{\ensuremath{A}\xspace}
\newcommand{\accSalt}{\ensuremath{Salt_{\mathrm{A}}}\xspace}
\newcommand{\accPasswordPolicy}{\ensuremath{PP_{\mathrm{A}}}\xspace}
\newcommand{\accData}{\ensuremath{Data_{\mathrm{A}}}\xspace}
\newcommand{\accDataDefinition}{\ensuremath{Data_{\mathrm{A}} = (Salt_{\mathrm{A}}, PP_{\mathrm{A}}, Username_{\mathrm{A}}, url_{\mathrm{A}})}\xspace}
\newcommand{\accID}{\ensuremath{ID_{\mathrm{A}}}\xspace}
\newcommand{\accIDDefinition}{\ensuremath{ID_{\mathrm{A}} = \mathrm{HMAC}(\accDataKeyMac, url_{\mathrm{A}})}\xspace}

\newcommand{\authK}{\ensuremath{K_{\mathrm{Auth}}}\xspace}

\newcommand{\authKBD}{\ensuremath{K_{\mathrm{Auth,BD}}}\xspace}
\newcommand{\authPK}{\ensuremath{PK_{\mathrm{Auth}}}\xspace}
\newcommand{\authPKUD}{\ensuremath{PK_{\mathrm{Auth,UD}}}\xspace}
\newcommand{\authPKBD}{\ensuremath{PK_{\mathrm{Auth,BD}}}\xspace}
\newcommand{\authPKBDPrime}{\ensuremath{PK_{\mathrm{Auth,BD'}}}\xspace}
\newcommand{\authPKPrimeBD}{\ensuremath{PK'_{\mathrm{Auth,BD}}}\xspace}
\newcommand{\authSK}{\ensuremath{SK_{\mathrm{Auth}}}\xspace}
\newcommand{\authSKUD}{\ensuremath{SK_{\mathrm{Auth,UD}}}\xspace}
\newcommand{\authSKBD}{\ensuremath{SK_{\mathrm{Auth,BD}}}\xspace}
\newcommand{\authToken}{\ensuremath{T_{\mathrm{Auth}}}\xspace}
\newcommand{\authTokenUD}{\ensuremath{T_{\mathrm{Auth,UD}}}\xspace}
\newcommand{\authTokenBD}{\ensuremath{T_{\mathrm{Auth,BD}}}\xspace}
\newcommand{\authTokenSSS}{\ensuremath{T_{\mathrm{Auth,SSS}}}\xspace}
\newcommand{\authCert}{\ensuremath{C_{\mathrm{Auth}}}\xspace}
\newcommand{\authKSSSGeneration}{\ensuremath{K_{\mathrm{Auth,SSS}} = \mathrm{KDF}(K_{\mathrm{Auth}}, url_{\mathrm{SSS}}) }\xspace}

\begin{document}

\title{Update-tolerant and Revocable\\ Password Backup (Extended Version)\thanks{Extended version of the paper that appeared in the
		proceedings of ACISP 2017.}}

\author{Moritz Horsch \and Johannes Braun \and Dominique Metz \and\ \\ Johannes Buchmann}

\institute{Technische Universit{\"a}t Darmstadt\\
	Hochschulstra{\ss}e 10\\
	64283 Darmstadt, Germany\\
	\path|{horsch,jbraun,metz,buchmann}@cdc.informatik.tu-darmstadt.de|}

\maketitle

\begin{abstract}
It is practically impossible for users to memorize a large portfolio of strong and individual passwords for their online accounts. A solution is to generate passwords randomly and store them. Yet, storing passwords instead of memorizing them bears the risk of loss, e.g., in situations where the device on which the passwords are stored is damaged, lost, or stolen. This makes the creation of backups of the passwords indispensable. However, placing such backups at secure locations to protect them as well from loss and unauthorized access and keeping them up-to-date at the same time is an unsolved problem in practice.

We present PASCO, a backup solution for passwords that solves this challenge. PASCO backups need not to be updated, even when the user's password portfolio is changed. PASCO backups can be revoked without having physical access to them. This prevents password leakage, even when a user loses control over a backup. Additionally, we show how to extend PASCO to enable a fully controllable emergency access. It allows a user to give someone else access to his passwords in urgent situations. We also present a security evaluation and an implementation of PASCO.
\end{abstract}

\section{Introduction}
\label{sec:intro}

Online accounts are mainly protected by passwords. Services can implement password-based authentication with little effort and operate it with a negligible cost per user \cite{DBLP:journals/ieeesp/HerleyO12}. Users are familiar with passwords and can use them across many platforms, devices, and applications \cite{DBLP:conf/sp/BonneauHOS12}. A password kept in user's mind appears to him to be always available, not threatened by loss like a key, and to be stored in the safest place on earth to which no one else has access and could steal it. Moreover, in emergency situations, a user can give a password to someone else to access an account on his behalf just by telling it. These perceived features of passwords are crucial for users \cite{DBLP:conf/icisc/KaroleSC10,DBLP:conf/soups/StobertB14,DBLP:journals/compsec/ZhaoY14}.

But, passwords pose a fundamental challenge to users: To resist the various known attacks against passwords \cite{DBLP:conf/sp/Bonneau12,DBLP:journals/corr/abs-1304-6584,DBLP:conf/sp/KelleyKMSVBCCL12,DBLP:conf/ccs/WeirACS10,DBLP:conf/sp/WeirAMG09}, users need to select a strong and different password for each account. These two security conditions require a large portfolio of strong and individual passwords to be memorized, which is practically impossible for users \cite{DBLP:conf/uss/FlorencioHO14,DBLP:conf/www/FlorencioH07,DBLP:conf/soups/GawF06}.

One solution is to create passwords randomly and store them on a user device, such as done by password managers \cite{OnePassword,dashlane,lastpass}. However, this approach bears the risk that the passwords get lost and has the drawback that the passwords might not be available on all user devices \cite{DBLP:conf/icisc/KaroleSC10}. Creating a backup of the passwords as well as copying them to all devices manually is burdensome and time-consuming, as it must be done after any change of the user’s password portfolio. This happens frequently as it includes new passwords and changing existing ones. Storing passwords in the cloud, for synchronization and backup, bears the risk of server compromises \cite{lastpasssn2}, security flaws \cite{DBLP:conf/esorics/GastiR12,DBLP:conf/uss/LiHAS14,DBLP:conf/uss/SilverJBCJ14,DBLP:conf/ccs/StockJ14,DBLP:conf/prisms/ZieglerRSTH14}, and governmental access \cite{cloudSec}. Although the passwords are encrypted, the security relies on a potentially weak, user-chosen memorable password. There are also hybrid solutions, that store some parts on a server and keep a strong invariable secret offline on the devices. However, loss protection for the offline secret must be in place meaning a secure and reliable backup solution is required. As well, the provision of emergency access for such hybrid solutions is still an open challenge, as the offline secret cannot be handed over easily or even worse, might not be available to the user himself in such a situation.

In this paper, we advance the \textit{PasswordLess Password Synchronization} (PALPAS) scheme \cite{DBLP:conf/IEEEares/HorschHB15} with a secure and usable backup solution called PASCO (\underline{PA}LPA\underline{S} RE\underline{CO}VERY). It allows users to recover their passwords in case their devices get lost. Additionally, our backup solution can be used to establish fully controllable emergency access to the passwords. Once a backup is created, it never needs to be updated even when the password portfolio changes. Therefore, it can be kept completely offline in secure, different, and physical isolated locations which minimizes the risk of compromise and loss. Furthermore, the backup solution has an built-in revocation mechanism, which allows the user to completely invalidate a backup if he loses control over it. The revocation mechanism works without having access to the backup itself and guarantees that no passwords can be leaked from it once revoked. 

To the best of our knowledge, PALPAS is the most secure hybrid password management solution currently available. It creates strong and individual passwords for accounts and makes them available on all user devices. PALPAS does not store passwords, neither on devices nor on servers. It generates them only when needed. This is done using a secret stored on all devices and some synchronization data stored on a server.

The combination of PALPAS and PASCO is the first full-fledged solution that ensures the confidentiality, availability, and recoverability of a user's password portfolio. Moreover, by integrating our backup revocation mechanism directly into PALPAS, its security can further be increased. This would enable the complete invalidation of PALPAS data stored on a user device, for example in case the device is stolen.

The paper is organized as follows: In Section \ref{sec:relatedWork} we summarize related work and we present the background about PALPAS in Section \ref{sec:palpas}. We describe PASCO in Section \ref{sec:pasco} and present its extension for emergency access in Section \ref{sec:pasco2}. A security evaluation follows in Section \ref{sec:security} and we present an implementation of PASCO in Section \ref{sec:implementation}. We conclude the paper in Section \ref{sec:conclusion}. Appendix \ref{app:recoverSSSData} provides an alternative approach to guarantee the availability of the data stored on the PALPAS synchronization server.

\section{Related Work}
\label{sec:relatedWork}

In 2007, Flor{\^{e}}ncio and Herley \cite{DBLP:conf/www/FlorencioH07} reported that users on average have 25 accounts. Due to the growth in online services, the number has strongly increased over the last ten years, which makes the memorization of secure passwords for all accounts practically impossible. Studies have shown that users typically cope with this challenge by selecting passwords that are easy to remember and reuse passwords across accounts \cite{DBLP:conf/ndss/DasBCBW14,DBLP:conf/soups/GawF06,DBLP:journals/cacm/IvesWS04,DBLP:conf/chi/UrBSBCC16,DBLP:conf/interact/ZezschwitzLH13,DBLP:conf/soups/WashRBW16}. This makes the passwords vulnerable to various attacks such as brute-force \cite{DBLP:conf/sp/KelleyKMSVBCCL12,DBLP:conf/ccs/WangZWYH16,DBLP:conf/sp/WeirAMG09}, dictionary \cite{DBLP:conf/sp/Bonneau12,DBLP:conf/ccs/WeirACS10}, and social engineering \cite{DBLP:journals/corr/abs-1304-6584}.

Beside many approaches to simplify the creation and memorization of passwords \cite{DBLP:conf/ndss/BlockiKCD15,DBLP:conf/uss/BonneauS14,DBLP:conf/chi/EgelmanSMBH13,DBLP:conf/uss/FlorencioHO14,DBLP:conf/www/HaldermanWF05,DBLP:journals/ijmms/HaqueWS14,DBLP:conf/ndss/KieselSL17,DBLP:conf/chi/ShayBCCFKMMSU15,DBLP:conf/chi/ShayKDHMSUBCC14}, storing passwords on user devices is the most common approach to solve the memorability problem. A prominent example are password managers \cite{OnePassword,dashlane,lastpass}. They store the user's passwords in a database, encrypted with a user-chosen master password. To synchronize the database between devices and prevent its loss, it is stored on a server. In emergency situations, a user can give someone else the master password, but then the person has access to all passwords. Moreover, a security breach at the server \cite{lastpasssn2} allows adversaries to steal the database and to perform offline brute-force attacks \cite{DBLP:conf/prisms/ZieglerRSTH14}. This can be mitigated by databases using honey encryption \cite{DBLP:conf/sp/ChatterjeeBJR15,DBLP:conf/ccs/JuelsR13}. But, their design is challenging \cite{DBLP:journals/tdsc/Erguler16,DBLP:conf/ccs/GollaBD16} and a backup concept does not exist yet.

Another approach is to only store data on servers that is independent from the passwords, as done by PALPAS \cite{DBLP:conf/IEEEares/HorschHB15}, which we use in this paper. For a general model of such schemes we refer to Al Maqbali et al.\ \cite{DBLP:conf/wistp/MaqbaliM16}. 

Storing the passwords on multiple user devices instead of a server does not serve as a proper backup concept because of the serious threat of malware. Already in 2010, ZeuS showed how to compromise multiple user devices even when they are not connected \cite{DBLP:conf/ccs/FeltFCHW11}. It infected the user's desktop computer and then tricked the user into installing the malware on his mobile devices. Moreover, password backups on external hard drives or network storage devices are also threatened by malware. For instance, CryptoLocker encrypts files on these devices and holds them hostage until the user pays a ransom \cite{DBLP:journals/corr/BhardwajSAS15,DBLP:conf/dimva/KharrazRBBK15}. Therefore, it is recommended to store backups on read-only memories such as CDs and put them on secure, different, and physical isolated locations \cite{fbiransomware,uscertransomware}. However, this is challenging for users, because their password portfolio changes frequently. We tackle these problems by using a secure element that is not required to be updated at any time.

\section{Background: PALPAS}
\label{sec:palpas}

Our work is based on PALPAS \cite{DBLP:conf/IEEEares/HorschHB15}. We start by explaining its functionality in Sections \ref{sec:palpas:pwgen} and \ref{sec:palpas:sss}. Then, we discuss the risk of password loss in Section \ref{sec:palpas:passloss}, for which we present a solution in Section \ref{sec:pasco}.

\subsection{Password generation}
\label{sec:palpas:pwgen}

\begin{figure}[b]
	\centering
	\includegraphics[width=.59\linewidth]{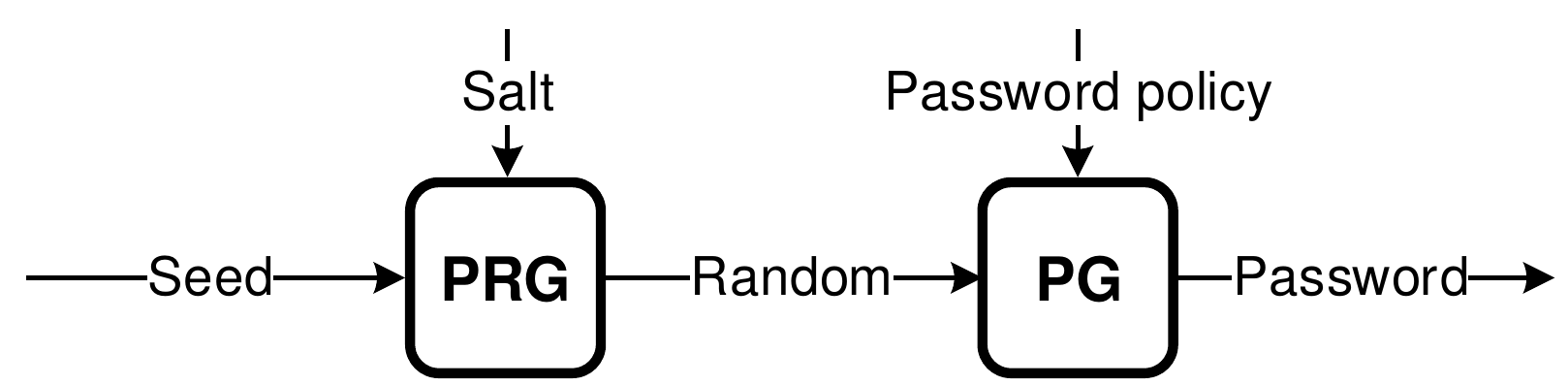}
	\caption{PALPAS password generation \cite{DBLP:conf/IEEEares/HorschHB15}.}
	\label{fig:Password_Generation}
\end{figure}

As illustrated in Figure \ref{fig:Password_Generation}, PALPAS generates a password for a user account in two steps: First, a (cryptographically secure) Pseudorandom Generator (PRG) generates a random value based on a \seed and a \salt. Second, a Password Generator (PG) derives a password from the random value and ensures that it complies with a password policy (\passwordpolicy). The PRG and the PG are deterministic. Thus, using the same \seed, \salt, and \passwordpolicy, the same password is generated.

The \seed is a randomly generated secret that is used for the generation of all passwords. It is created when a user uses PALPAS for the first time and it does not change over time.

The \salt is also a random value, but it differs for each account so that individual passwords for different accounts are created. Changing the \salt for an existing account allows to create a new password, e.g., during a regular password change. An initial salt is generated when PALPAS initially creates the password for an account, e.g., during account creation.

The \passwordpolicy specifies the password requirements of a service such as the password length and the allowed characters. By this, PALPAS ensures that the randomly created password is actually accepted by the service. The \passwordpolicy is created when a service is used for the first time. This can be done manually by the user or the \passwordpolicy is retrieved from a central service as presented by Horsch et al.\ \cite{DBLP:conf/acisp/HorschS0B16}. During password change for an account (by changing the salt) it might also be necessary to update the \passwordpolicy in order to comply with the recent password requirements of a service.

\subsection{Password synchronization}
\label{sec:palpas:sss}

PALPAS creates a password portfolio of strong and individual passwords for the user's accounts using a fixed \seed and an individual \salt and a \passwordpolicy for each service, respectively. To enable the computation of the passwords on different devices, the \seed is shared by all user devices and the salts and policies are synchronized between them through a server, namely the Salt Synchronization Service (SSS).

When creating new or changing passwords, the corresponding salts and policies are added or updated at the SSS. Each time PALPAS recomputes a password, the corresponding \salt and \passwordpolicy for the account is retrieved from the SSS. In this way, any changes of the user's password portfolio are immediately available on all of his devices.

Beside the salts and policies also the usernames of the user's accounts are synchronized between the devices through the SSS. For each user account \acc an account data object \accDataDefinition is stored at the SSS, where $url_{\mathrm{A}}$ is the service's URL. Each \accData is encrypted with a key \accDataKeyEnc by PALPAS before it is transferred to the SSS. To retrieve only the \accData for a specific account and not all of them, each \accData is associated with an identifier \accID. It is generated by PALPAS by \accIDDefinition, where $url_{\mathrm{A}}$ is the service's URL. Both keys, \accDataKeyEnc and \accDataKeyMac, are derived from a key \accDataKey which is randomly created when a user uses PALPAS for the first time.

Moreover, the user's account at the SSS is protected from unauthorized access. Each user device has an individual key pair \authK for authentication, consisting of a private key \authSK and public key \authPK. The key pair is randomly created by the user device when PALPAS is used for the first time and an account at the SSS is created. \authSK is stored on the device and \authPK is transferred to the SSS.

To register multiple devices for an account at the SSS, an already registered device needs to request an authentication token \authToken. The new device creates its own key pair \authK and uses \authToken to register its \authPK at the SSS. \authToken is randomly created by the SSS and has a limited validity. All communication with the SSS is sent through a mutually authenticated and secure channel, e.g. using TLS.

To set up PALPAS on all of his devices, a user needs to transfer the PALPAS secret \secretDef to each device and register the device at the SSS using an authentication token. This has to be done only once. The data transfer can be easily achieved by a file transfer or a QR code.

\subsection{Password loss}
\label{sec:palpas:passloss}

The \seed, the salts, and the password policies are crucial for password generation. To retrieve the account data from the SSS, \authSK and \accDataKey are required. The availability of these five pieces of data must be guaranteed. Otherwise, the user's password portfolio is lost.

The account data is stored at the SSS. Their availability has to be guaranteed by the SSS. We assume that the provider of the SSS implements proper measures to restore the data at any time. In case of cloud storage providers, this can be realized with standard means like redundant hardware and different data centers. As an alternative, we describe a user-side solution using multiple SSSs in Appendix \ref{app:recoverSSSData}.

The PALPAS secret \secretDef and the individual \authSK are exclusively stored on user devices and thus are at high risk of loss. Typical cases are lost, stolen, or damaged devices as well as malware.

\section{PASCO}
\label{sec:pasco}

We now present PASCO (\underline{PA}LPA\underline{S} RE\underline{CO}VERY), a secure and usable backup solution for PALPAS. It ensures that users never lose their password portfolio by providing recoverability of the essential PALPAS data that is stored on the user device.
\clearpage
PASCO uses a separate backup device (BD) to store a backup of the PALPAS data. We consider the BD to be a tamper resistant device that provides secure storage, user authentication, and basic cryptographic algorithms\footnote{In Section \ref{sec:implementation} we present a implementation of PASCO using a smart card as a BD.}. The BD stores the PALPAS secret \secretDef encrypted by a one-time-pad (OTP) \cite{DBLP:journals/cryptologia/Bellovin11,miller1882telegraphic}. Furthermore, it has its own authentication key pair \authKBD for the SSS. The BD is protected by a user-chosen PIN. To prevent guessing attacks, it has a retry counter for the PIN. After five wrong PIN entries the BD erases all stored data.

We explain the creation of the update-tolerant backup in Section \ref{sec:pasco:creation} and how to restore the PALPAS data from a backup in Section \ref{sec:pasco:restore}. In Section \ref{sec:pasco:revocation}, we describe the built-in revocation mechanism for BDs, which protects the user's password portfolio from leakage, even when the user loses control over a BD.

\subsection{Creating a backup}
\label{sec:pasco:creation}

\begin{figure}[h]
	\centering
	\includegraphics[width=.9\linewidth]{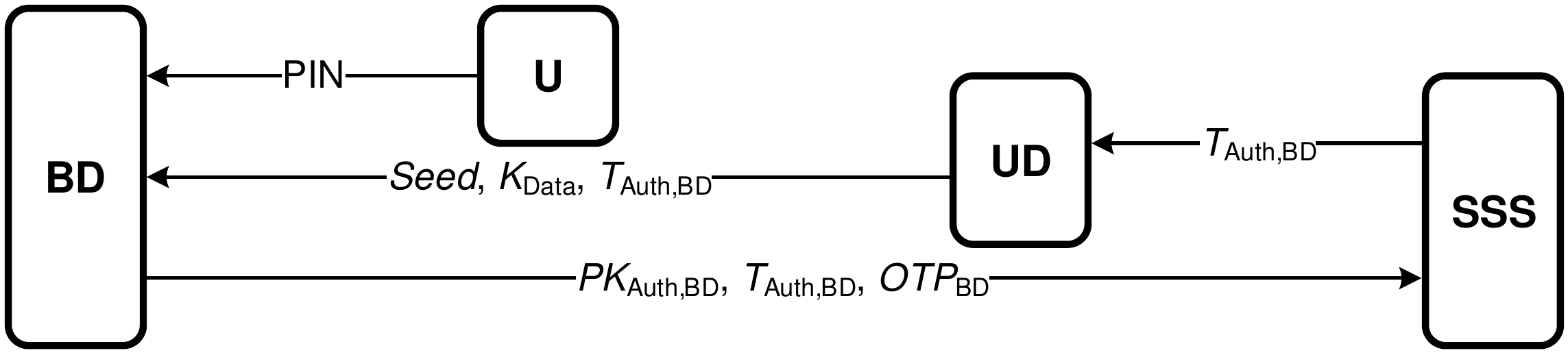}
	\caption{Data flow of the PASCO backup procedure.}
	\label{fig:nameit:backupprocedure}
\end{figure}

We assume that the user already uses PALPAS and has registered a device at the SSS. The procedure to create a PASCO backup is described in the following. The data flow is illustrated in Figure \ref{fig:nameit:backupprocedure}.

\begin{enumerate}
	\item The user (U) initializes the BD with a PIN.
	\item The user device (UD) connects to the SSS, authenticates itself with its \authSKUD, and requests \authTokenBD.
	\item The UD sends \secretDef and \authTokenBD to the BD.
	\item The BD randomly samples a one-time-pad key \secretOTP and computes \secretOTPEnc. Then, it generates a key pair \authK. \authSKBD is stored at the BD and \authPKBD, \authTokenBD, and \secretOTP is send to the SSS. The SSS verifies \authTokenBD and stores \authPKBD and \secretOTP. Finally, BD stores \secretBD and deletes \secretOTP and \secret.
\end{enumerate}

An update of the BD is not necessary, even when the user's password portfolio is changed. Changing, adding, or deleting passwords only requires to update, store, or delete the related \accData at the SSS. This is already an integral part of PALPAS. As the SSS provides availability of the account data, PASCO itself does not need to take care of it.

\subsection{Restoring data from a backup}
\label{sec:pasco:restore}

\begin{figure}[b]
	\centering
	\includegraphics[width=.9\linewidth]{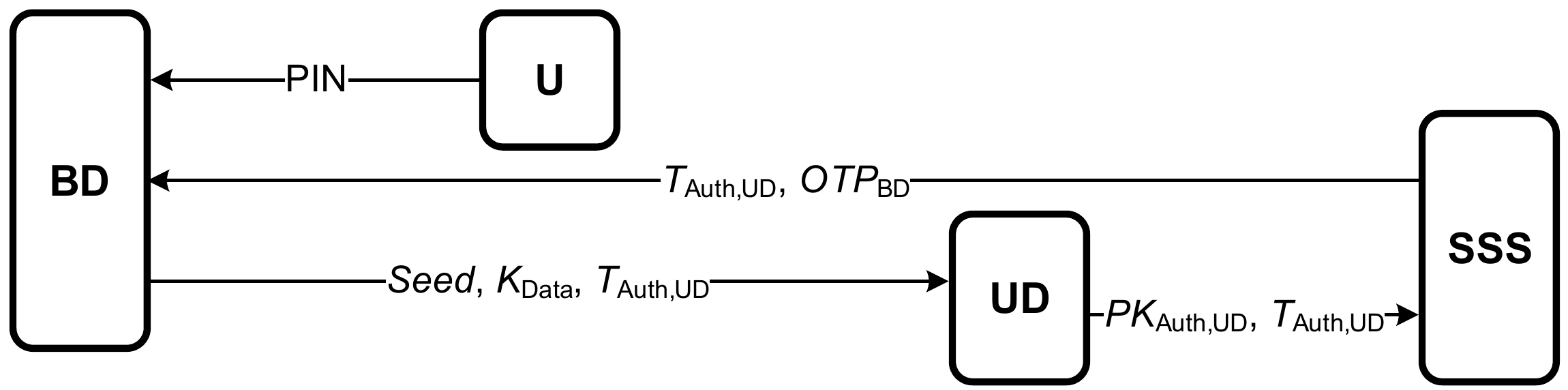}
	\caption{Data flow of the PASCO restore procedure.}
	\label{fig:nameit:restoreprocedure}
\end{figure}

To restore the PALPAS data on a device, the user needs to have the BD, the corresponding PIN, and a user device with PALPAS. The restoring works as follows. The data flow is illustrated in Figure \ref{fig:nameit:restoreprocedure}.

\begin{enumerate}
	\item The user (U) authenticates himself to the BD with his PIN. 
	\item The BD contacts the SSS, authenticates itself with its \authSKBD, and requests \authTokenUD and \secretOTP.
	\item The BD computes \secretOTPDec. Then, it transfers the \seed, \accDataKey, and \authTokenUD to the UD. The UD stores the \seed and \accDataKey.
	\item The UD creates a key pair \authK. \authSKUD is stored at the UD and \authPKUD and \authTokenUD is send to the SSS. The SSS verifies \authTokenUD and stores \authPKUD. UD has now access to the user's account at the SSS and can retrieve the account data for generating the passwords.
\end{enumerate}

\subsection{Revoking a backup}
\label{sec:pasco:revocation}

All existing BDs are registered at the SSS with their individual \authPKBD and \secretOTP. To revoke a BD a user deletes the related \authPKBD and \secretOTP at the SSS. Now the BD can no longer retrieve any account data nor request new authentication tokens. Moreover, the deletion of the \secretOTP invalidates \secretBD and it is impossible to recover the PALPAS secret \secret from the BD. Thus, the BD is useless and the passwords cannot get leaked (see also Section \ref{sec:security}). Our revocation solution based on a OTP can also be integrated into PALPAS to securely revoke stolen user devices.

\section{PASCO backups with emergency access}
\label{sec:pasco2}

Storing passwords on devices relieves users from memorizing their passwords. However, not knowing the passwords makes it impossible to give someone else access to an account in urgent or emergency situations.

We now describe how a user can allow someone else to use a BD to access his accounts. In Section \ref{sec:pasco2:extension}, we extend PASCO to enable a fully controllable emergency access. In Section \ref{sec:pasco2:creation}, we describe the creation and management of such a BD by the user. Particularly, we describe how to control which accounts can be accessed with a specific BD. In Section \ref{sec:pasco2:access}, the procedure for an emergency access is explained.

\subsection{Extending PASCO}
\label{sec:pasco2:extension}

In addition to storing the PALPAS data, the BD now implements the PALPAS password generation procedure (cf.\ Section \ref{sec:palpas}). Furthermore, the SSS is equipped with a fine granular access control system for the account data. For each \authPK, a user can specify different access rules to the account data. While, one \authPKBD may have access to all data, another \authPKBDPrime can only access the data for the user's mail account.

\subsection{Creating and managing a backup  with emergency access}
\label{sec:pasco2:creation}

The procedure for creating a BD with emergency access is nearly the same as described in Section \ref{sec:pasco:creation}. It only differs in the second step, where the UD requests \authTokenBD. The request is supplemented by an access control list (ACL), which is basically a list of account data identifiers (cf. Section \ref{sec:palpas:sss}). The \authPKBD registered using \authTokenBD is later only granted access to the account data defined by the ACL. The ACL for each \authPK can be modified at any time. For instance, the user can enable the access to his social media account during vacation and disable it afterwards. Both can be done without having physical access to a BD. Moreover, a user can also revoke a BD to invalidate it completely (cf.\ Section \ref{sec:pasco:revocation}).

The BD can simultaneously act as a backup and password generation device. Thus, depositing a single BD at a friend's place is sufficient. To provide both features, the BD is equipped with multiple authentication keys. One \authPKBD is allowed to request an \authTokenUD as needed for the restoring procedure (cf.\ Section \ref{sec:pasco:restore}). Another \authPKPrimeBD can only retrieve certain account data and is used for the emergency access. To equip a BD with multiple authentication keys, the aforementioned creation procedure is performed multiple times with a different \authToken, \authPK, ACL, and in particular a different PIN. Depending on the PIN, the BD uses the corresponding \authPKBD for the authentication at the SSS. In accordance with the ACL associated to \authPKBD, the SSS allows to request an \authToken or only certain account data, e.g.\ \accData for the user's mail account.

\subsection{Accessing a backup in case of an emergency}
\label{sec:pasco2:access}

\begin{figure}[h]
	\centering
	\includegraphics[width=.9\linewidth]{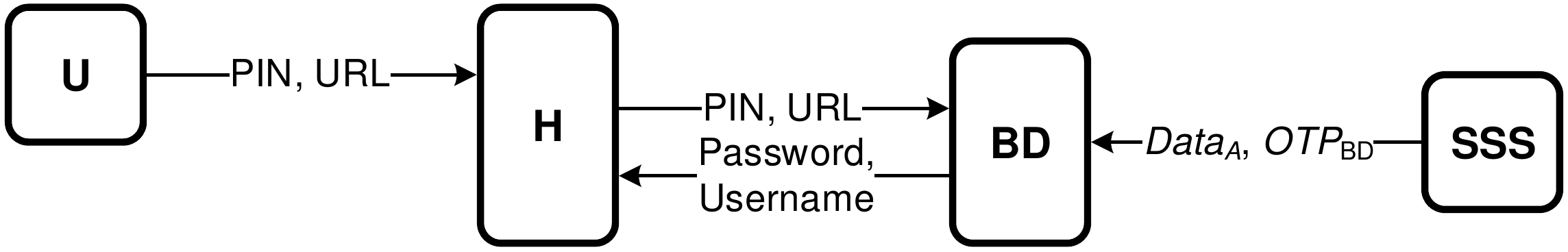}
	\caption{Data flow of the PASCO emergency access procedure.}
	\label{fig:nameit:accessprocedure}
\end{figure}

We envision that a user has deposited a PASCO backup at a friend's place. Allowing the friend (i.e.\ BD holder) to create the user's password for an account works as follows. The data flow is depicted in Figure \ref{fig:nameit:accessprocedure}.

\begin{enumerate}
	\item The user (U) tells the BD holder (H) the emergency PIN of the BD and the URL of the service where H should access the user's account.
	\item H uses the PIN to authenticate himself to the BD and transfers the URL of the service to the BD.
	\item The BD connects to the SSS and authenticates itself with \authSKBD. It calculates \accID for the URL and requests the corresponding \accData.
	\item The SSS checks the ACL for \authPKBD and, if the access is allowed, returns \accData and \secretOTP.
	\item The BD computes \secretOTPDec and then decrypts \accData with \accDataKeyEnc to obtain \accSalt and \accPasswordPolicy. Finally, it generates the password using the \seed, \accSalt, and \accPasswordPolicy (cf.\ Section \ref{sec:palpas}).
	\item The BD deletes \secret and hands the password and username over to H. H can now browse the service and log in to the user's account.
\end{enumerate}

\section{Security evaluation}
\label{sec:security}

A security evaluation of PALPAS can be found in \cite{DBLP:conf/IEEEares/HorschHB15}. In this section, we focus on the security implications of the new features and additional data stored at the SSS as introduced by PASCO. We consider the following three scenarios: The theft/loss of a BD, a person with emergency access becoming untrustworthy, and a security breach at the SSS. Throughout the evaluation, we assume that the BD was securely created.

\subsubsection*{Theft or loss of a BD}
In case of theft or loss, the BD is protected with a PIN and a retry counter, which prevents a brute-force attack. As the BD stores all data in a secure storage this provides a first line of defense.

Now we consider a stronger adversary, being capable to circumvent the PIN protection and extract the data from a stolen BD, e.g., by timing or side-channel attacks \cite{DBLP:conf/esmart/RomerS01}. The adversary obtains \secretBD and \authSKBD. However, \secretBD is one-time-pad encrypted and therefore statistically independent from \secret as long as the adversary does not obtain the corresponding key \secretOTP. Given BD has been revoked in the meantime, \authSKBD has been invalidated and the adversary cannot request any data or an authentication token from the SSS. Moreover, \secretOTP has been deleted and it is impossible to reconstruct the PALPAS secret \secret from \secretBD. Thus, the PIN only needs to protect the BD until it is revoked and gets useless. There is no need for the user to change his passwords, the \seed, or \accDataKey.

\subsubsection*{Emergency access}
As long as the emergency PIN is not revealed to the BD holder, the scenario is the same as in case of theft/loss. Revoking the BD makes it useless. In case the PIN was revealed, the user can change the BD's access rights for the account data or revoke the BD completely. A change of the revealed passwords ends access to the user's accounts.

\subsubsection*{Security breach at the SSS}
Deviating from the original PALPAS, the SSS additionally stores the OTPs used to encrypt the PALPAS secret on the BDs. The OTPs are randomly sampled and statistically independent from \secret. This means, an adversary learning the OTPs has no advantage over an adversary analyzed for the original PALPAS scheme (cf. \cite{DBLP:conf/IEEEares/HorschHB15}). Thus, PASCO does not decrease the security of PALPAS concerning a security breach at the SSS.

\section{Implementation}
\label{sec:implementation}

We now present an implementation of PASCO and in particular show the realization of the BD using a smart card.

We used a Java card (NXP J3D081) and developed a Java Card Applet (Classic Platform) that implements the backup and emergency feature. The PALPAS data is stored in the secure memory of the card and is protected by a PIN. The SSS was implemented as a Java Web Service and a PALPAS/PASCO client as a Java desktop application. Due to the restrictions of smart cards, the application and communication flows of our implementation slightly differ from the conceptual description. Nevertheless, our realization fulfills the pursued security features. Namely, it securely stores the PALPAS data and protects it from unauthorized access and other threats like malware.

\paragraph{Authentication and communication}
\label{sec:implementation:a}

Authentication at the SSS is realized using TLS with mutual authentication. \authK is a RSA key pair along with a corresponding X.509 \cite{RFC5280} certificate \authCert issued by the SSS. To initially create an account at the SSS, a device creates \authK and a Certificate Signing Request (CSR) \cite{RFC2986}. The CSR is sent to the SSS which in turn issues \authCert. Whenever another device is added to an account, a CSR (containing \authPK of the new device), \authToken and, in case a BD is registered, a \secretOTP is sent to the SSS. \authToken is realized as a random 32 bytes value with a validity period of 5 minutes.

The smart card and the SSS cannot communicate directly, because of deviating communication protocols. Therefore, our PALPAS/PASCO application acts as a proxy. During the restore and emergency procedure it establishes a TLS channel to the SSS and a local connection to the BD. The application forwards all messages between SSS and BD to hand over \secretOTP and to perform mutual authentication using \authCert and \authSK stored on the card. \authSK does never leave the BD.

\paragraph{Password generation}
\label{sec:implementation:b}

Other than in the conceptual description of the emergency access function, the password generation is jointly done by the BD and the PALPAS/PASCO application. The first part, the generation of the random (cf.\ Figure \ref{fig:Password_Generation}), is done by the BD. After recovering the \seed from  \secretBD and \secretOTP and decrypting the \salt and the \passwordpolicy, it computes the random, and hands it together with the \passwordpolicy over to the application. The second part is done by the application. It takes the random and the \passwordpolicy and computes the password. The joint approach was necessary, because the processing of the XML-encoded password policies as used by PALPAS could not be realized on the card. Yet, this does not pose any security issues, because the \seed and \accDataKey never leave the smart card.

\section{Conclusion}
\label{sec:conclusion}

Storing passwords on user devices is the most suitable approach to realize both strong and individual passwords for user accounts. However, to serve as a secure and usable solution, the confidentiality, availability, and recoverability of the stored passwords must be ensured. The combination of PALPAS and our PASCO provides the first solution that fulfills all these three requirements. With the implementation we have shown that PASCO can be realized in practice with an off-the-shelf smart card.

We have also presented a revocation mechanism that could additionally be integrated into PALPAS to enable the secure revocation of user devices. 
With the emergency access, we address a major concern of users regarding the storage of passwords. Moreover, this function can be used to generate passwords in general. Storing the PALPAS data on a PIN-protected smart card within its protected memory provides much more security compared to encrypted storage on a user device. 
In particular this is a major advantage in the mobile environment, which we will focus in our future work. The smart card that we used has a contactless interface and thus it is capable to communicate with NFC-enabled mobile devices. Using the smart card as a password generator allows users to literally have their passwords in their wallet. With the two key features of PASCO, users need not update the card when their password portfolio is changed and they are able to revoke it in case of loss at any time.

\section*{Acknowledgment}
This work has been cofunded by the DFG as part of project S6 within the CRC 1119 CROSSING.

\bibliographystyle{abbrv}
\bibliography{references}

\appendix

\section{Recoverability of the SSS-side PALPAS data}
\label{app:recoverSSSData}

\begin{figure}
	\centering
	\begin{subfigure}[h]{0.5\textwidth}
		\centering
		\includegraphics[width=1\linewidth]{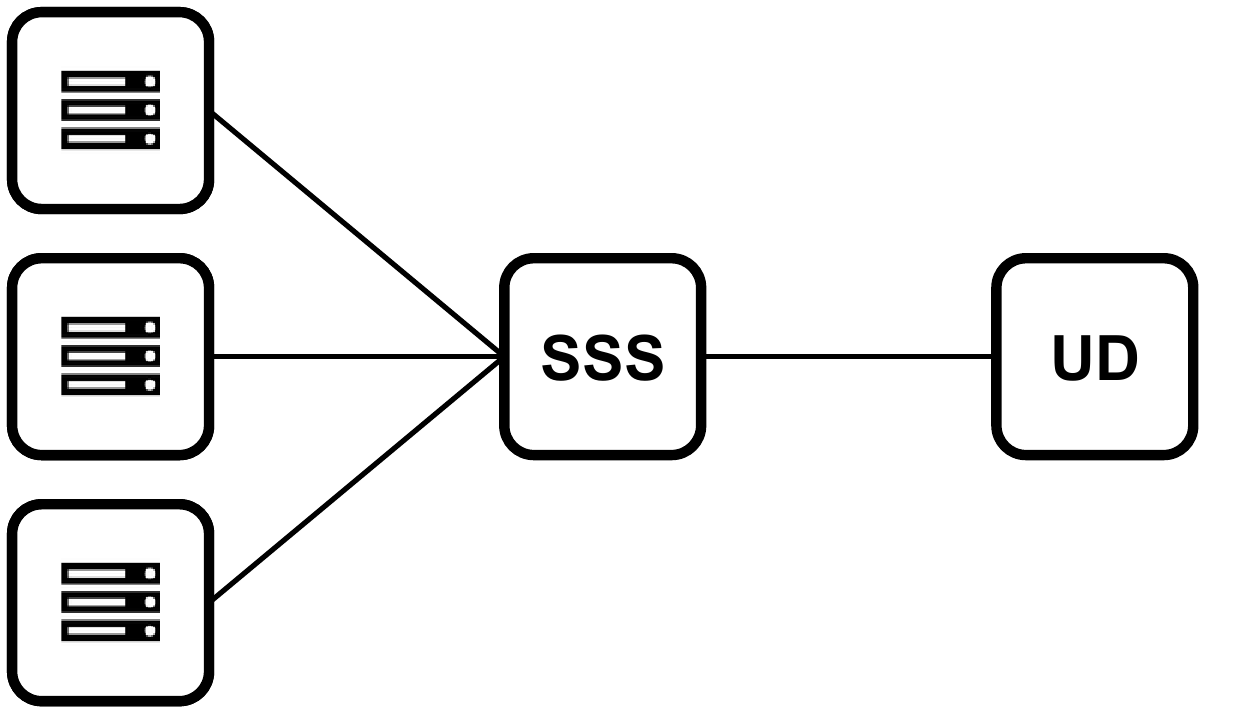}
		\caption{SSS-side solution}
		\label{fig:nameit:sss1}
	\end{subfigure}
	\hspace{.5cm}
	\begin{subfigure}[h]{0.35\textwidth}
		\centering
		\includegraphics[width=1\linewidth]{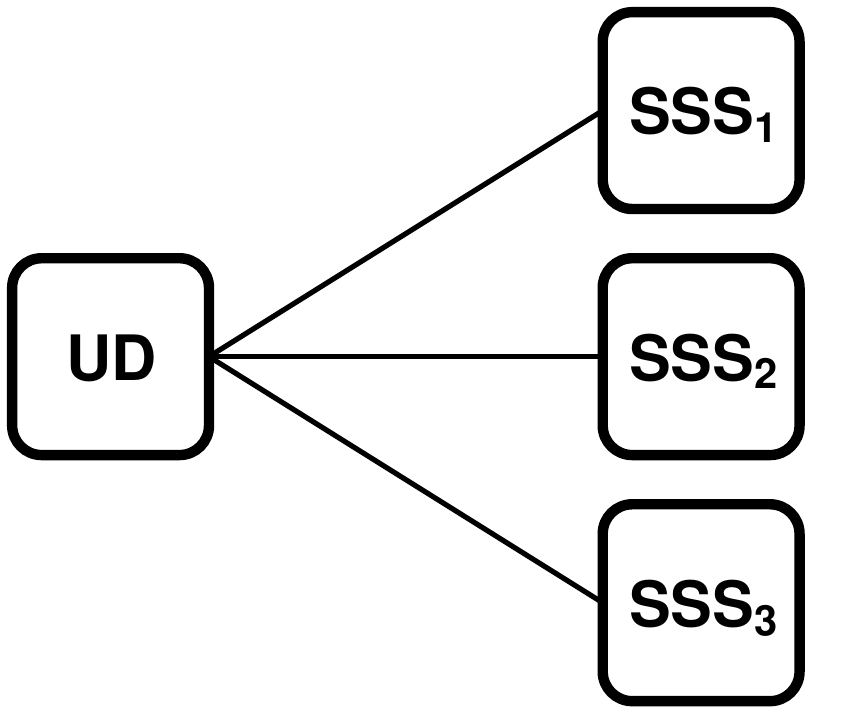}
		\caption{User-side solution}
		\label{fig:nameit:sss2}
	\end{subfigure}
	\caption{Solutions for the recoverability of the SSS-side PALPAS data.}
	\label{fig:nameit:sss}
\end{figure}

PASCO provides recoverability of the PALPAS data stored on the user device. In general, we assume that the recoverability of the account data stored on the SSS is guaranteed by the SSS provider. An SSS provider can realize this for example by using redundant hardware and multiple data centers as depicted in Figure \ref{fig:nameit:sss1}. 

However, for users that do not want to rely on this assumption, we now describe an alternative approach to protect the server-side data from loss. The basic approach is to redundantly store the account data on multiple SSSs to mitigate the risk of loss, as illustrated in Figure \ref{fig:nameit:sss2}. Besides loss protection, this also mitigates unavailability during a potential temporary outage or maintenance of an SSS.

Storing the data on multiple SSSs can be realized with minor adaption of the PALPAS client. Instead of storing the data on a single SSS, a user device mirrors the account data to one or more additional SSSs.The communication protocols and interfaces of the SSS providers need not to be changed. The additional network traffic is negligible, because the account data only make up for a few kilobytes. To efficiently create individual authentication and encryption keys for different SSSs, we propose a new scheme to generate these keys. As well, we show how a one-time-pad key can me masked in order to provide privacy protection.

In the current version of PALPAS \accDataKey and \authK are randomly generated. Instead of applying these keys directly, we propose to use them as input for a Key Derivation Function (KDF) to create SSS-specific encryption and authentication keys, respectively. In case PALPAS uses the OTP-based revocation mechanism introduced by PASCO and when PASCO is used with multiple SSSs, we propose to mask the OTP keys before storing them on the SSSs. This approach has two advantages: First, it protects the user's privacy because keys are not reused. Otherwise, collaborating SSS providers could identify users by comparing the authentication or OTP keys. Second, it does not require to store a multitude of keys so it can be realized on a smart card, where the storage capacity is very limited. Only a small random bit string needs to be additionally stored. In detail, the generation of SSS-specific encryption and authentication keys and the OTP masking work as follows.

\paragraph{SSS-specific encryption keys}
When a user uses PALPAS for the first time \accDataKey is randomly created. But, PALPAS does not use the key directly, it creates an SSS-specific key \accDataKeySSSGeneration on demand for each SSS, where $url_{\mathrm{SSS}}$ is the URL of the SSS. From \accDataKeySSS an encryption key \accDataKeyEncSSS and a message authentication key \accDataKeyMacSSS can be derived as in the original PALPAS. For setting up PALPAS on another device, the user just transfers \accDataKey to the new device as done in the original version of PALPAS. The URLs of the SSSs can be included in this transfer or manually entered.

\paragraph{SSS-specific authentication keys}
Each device still randomly creates its \authK, but does not use it directly for authentication. Instead, it creates an SSS-specific authentication \authKSSSGeneration key for each SSS, where $url_{\mathrm{SSS}}$ is the URL of the SSS. Note that in case of setting up PALPAS on a new device, the registered device needs to request a \authTokenSSS from each SSS. The tokens can be transferred together with the URLs of the SSSs, the \seed, and \accDataKey by a file transfer or a QR code.

\paragraph{SSS-specific OTP key masking}
When setting up a new device, PALPAS samples a random bitmask \secretOTPMask in addition to the one-time-pad key \secretOTPUD that is used to encrypt the PALPAS secret on the device. Then, it creates a SSS-specific bitmask \secretOTPMaskSSSGeneration, where $url_{\mathrm{SSS}}$ is the URL of the SSS and PRG a secure pseudorandom generator. Note that the URL of the SSS must be available to the devices anyway, because it is required for the registration/setup procedure at the SSS. Instead of \secretOTPUD  (cf. Step 4, Section \ref{sec:pasco:creation}), the UD sends \secretOTPSSSGeneration to the SSS. \secretOTPMask is stored on the user device while \secretOTPMaskSSS is recomputed when required to recover \secretOTPUD from \secretOTPSSS. With respect to PASCO, the procedure is the same to create a BD supporting multiple SSSs.

\end{document}